\documentclass[intlimits,twoside,a4paper]{article}

 \usepackage{graphicx}        
\usepackage{epstopdf}

\usepackage[cp1251]{inputenc}

\usepackage{tikz}
\usetikzlibrary{automata, arrows}

\usepackage[eqsecnum]{cmpj3}

\usepackage{bm}

\issue{2020}{23}{1}{13001}
\doinumber{10.5488/CMP.23.13001}

\title[Master-slave synchronization of Bose-Einstein condensate in 1D tilted bichromatical optical lattice]%
{Master-slave synchronization of Bose-Einstein condensate in 1D tilted bichromatical optical lattice%
}
\author[E. Tosyali, F. Aydogmus]{E. Tosyali\refaddr{label1},
        F. Aydogmus\refaddr{label2}}
\addresses{
\addr{label1} Istanbul Bilgi University, Vocational School of Health Services,  34387 Istanbul/Sisli, Turkey
\addr{label2} Istanbul University, Department of Physics, 34134 Istanbul, Turkey}

\date{Received	September 7, 2019, in final form September 26, 2019}

\begin{document}

\maketitle

\begin{abstract}
This paper investigates the synchronization of chaotic behavior in a model of Bose-Einstein condensate (BEC) held in a 1D tilted bichromatical optical lattice potential by using the active control technique. The synchronization is presented in the master-slave configuration which implies that the master system evolves freely and drives the dynamics of the slave system. Also the numerical simulations are given to indicate the practicability and the effectiveness of the used controllers.
\end{abstract}

\section{Introduction}
Chaos is a well known nonlinear phenomenon in many scientific disciplines, ranging from physics to engineering. A chaotic system has unpredictable and complex behaviors. Especially, there has been an increasing interest in the study of chaos synchronization. Chaos synchronization has great potential in physics, chemical and biological systems, secure communications, power electrical systems and so on \cite{Work70,Work71,Work72,Work73,Work74,Work75,Work76,Work77}. Carroll and Pecora first introduced the idea of synchronizing two chaotic systems in 1990~\cite{Work70}. Due to the connection between control and synchronization, numerous control methods are used to synchronize the chaotic systems. Among these methods, active control has been widely accepted as one of the efficient methods for chaos synchronization \cite{Work78,Work79}. Especially, it is very attractive in studying the control problems of the Bose-Einstein condensate (BEC) system due to its rich dynamics. The BEC system in an optical lattice shows many good properties as a typical nonlinear system \cite{Work1,Work2}. It is well known that the mean-field theory is a successful theory commonly used to describe the BEC. This approach can quite accurately describe the static and dynamical properties of BEC. The relevant model is a variant of the famous nonlinear Schr\"{o}dinger equation, the so-called Gross Pitaevskii (GP) equation~\cite{Work8,Work27}. 

In this paper, first we introduce the GP equation with 1D tilted bichromatical optical lattice potential. The differential equations system derived from GP equation is obtained by the use of a particular solution similar to solve the nonlinear Schr\"{o}dinger equation. Then, we focus on the synchronization of BEC system via an active control technique in the master-slave scheme. In the master-slave scheme, the behavior of the slave system is controlled with the operation of a master system, i.e., the master-slave configuration means that the master system evolves freely and drives the dynamics of the slave system \cite{Work80}. The controller provides that the states of the controlled chaotic slave system exponentially synchronize with the state of the master system.  Furthermore, numerical simulations for synchronization of the BEC system are performed to verify the results. In figure~\ref{fig1}, the schematic representation of the master-slave configuration is shown.

\section{Active control and synchronization}

Active control method is one of the most frequently used approaches for synchronization between two chaotic systems in  the master-slave scheme \cite{Work84}. If we consider a master system
\begin{equation}
\frac{\rd a}{\rd x}={\bf{A}} a+g(a)\,,
\label{eqn1.1}
\end{equation}
where $a$ state vector and ${\bf{A}}$ is a constant system matrix, and $g(a)$ is nonlinear function. And the slave system
\begin{equation}
\frac{\rd b}{\rd x}={\bf{B}}b+f(b)+u(x)\,,
\label{eqn1.2}
\end{equation}
where $b$ is state vector, ${\bf{B}}$ is a constant system matrix, and $f(b)$ is a nonlinear function, and $u(x)$ is an active control function.
 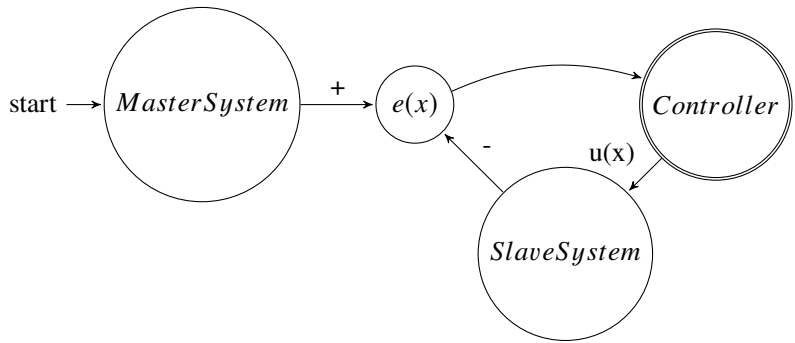
\begin{figure}[!b]
\begin{tikzpicture}[>=stealth',shorten >=1pt,auto,node distance=2.8cm]
  \node[initial,state] (q1)      {$Master System$};
  \node[state]         (q2) [right of=q1]  {$e(x)$};
  \node[state] (q4) [below right of=q2] {$Slave System$};
  \node[state, accepting]         (q3) [above right of=q4] {$Controller$};

  \path[->]          (q1)  edge                 node {+} (q2);
  \path[->]          (q2)  edge   [bend left=20]   node {} (q3);

  \path[->]          (q4)  edge                 node [swap] {-} (q2);
  \path[->]          (q3)  edge                 node [swap] {u(x)} (q4);
\end{tikzpicture}
\caption{The master-slave synchronization configuration.}
\label{fig1}
\end{figure}

In the master-slave synchronization configuration, the error state is obtained from $e=b-a$. Thus, the error dynamics is written as follows: 
\begin{equation}
\frac{\rd e}{\rd x}=\frac{\rd b}{\rd x}-\frac{\rd a}{\rd x}={\bf{C}}e+G(a,b)+u(x)\,,
\label{eqn1.3}
\end{equation}
where ${\bf{C}}={\bf{\bar{B}}}-{\bf{\bar{A}}}$ are the common parts of the system matrices, the non-common parts and nonlinear functions are gathered in $G(a,b)$ as follows:
\begin{equation}
G(a,b)=f(a)-g(b)+(B-\bar{B})b-(A-\bar{A})a\,,
\label{eqn1.4}
\end{equation}
and $u(x)$ is the controller matrix. The feedback from the controller $u(t)$ is designed to get the error $e$ to decay to zero. Thus, active controller should eliminate nonlinear terms and non-common parts. We define the active control function
\begin{equation}
u(x)=-G(a,b)+\tau(x)\,,
\label{eqn1.5}
\end{equation}
where $\tau(x)=-{\bf{K}} e$ is a linear controller and $\bf{K}$ is a linear gain matrix. After substitution of equation~(\ref{eqn1.5})  into~(\ref{eqn1.3}) we get
\begin{equation}
\dot{e}={\bf{C}}e+\tau(x)\,,
\label{eqn1.6}
\end{equation}
we replace $\tau(x)$ in to equation~(\ref{eqn1.6})
\begin{equation}
\dot{e}=({\bf{C}}-{\bf{K}})e
\label{eqn1.7}
\end{equation}

 If the eigenvalues of matrix ${\bf{C}}-{\bf{K}}$ are negative real or complex with negative parts, the system is synchronized and asymptotically stable at the origin. The controller manipulates the slave system and the controlled chaotic slave system exponentially synchronizes with the state of the master system as shown in figure~\ref{fig1}.

\section{The Gross-Pitaevskii equation}

In the ultra-cold temperature regime, the nonlinear Schr\"{o}dinger equation, known as the GP equation~\cite{Work27,Work9}, models quite accurately the static and dynamical properties of BEC with the macroscopic wave function $\Psi=\Psi(x,t)$. There are many different methods of reducing the original three-dimensional generalized GP equation to quasi-one dimensional GP equation. We consider a BEC described with quasi-1D cubic nonlinear GP equation which is used in the regime of shock waves~\cite{Work81, Work82, Work82a, Work83, Work15}, which is in the form of
\begin{equation}
\ri\hbar \frac{\partial }{\partial t}\Psi \left( x,t\right) =-\frac{\hbar ^{2}%
}{2m}\frac{\partial ^{2}}{\partial x^{2}}\Psi \left( x,t\right) +\left[
V_\text{ext}\left( x\right) +g_\text{1D}^{\prime}\left\vert \Psi \left( x,t\right) \right\vert
^{2}\right] \Psi \left(x,t\right),  \label{eqn2.1}
\end{equation}
where $V_\text{ext}$ ia an external potential confining the BEC, $m$ is the mass of the atoms of the condensate, $g_\text{1D}^{\prime}$ describes the quasi-1D interaction between the atoms in the condensate and is given by
 \begin{equation} 
 g_\text{1D}^{\prime}=\frac{m \omega_{r} g_\text{3D}}{2\piup a_{r}^{2}}=2a_{s} \hbar \omega_{r}\,, \nonumber  
 \end{equation}
where $g_\text{3D}=4\piup a_{s} \hbar^{2}/m$ is the atom-atom ineraction and proportional to the $a_{s}$. $a_{s}$ is the $s$-wave scattering length between the atoms. The $s$-wave scattering length is positive or negative for repulsive and attractive interactions, respectively. In our case, it is negative. $w_{r}$ is ground state of a harmonic frequency of the oscillator. Taking an experimentally suitable frequency $\omega_{0}$ as the units of frequencies $\omega_{1},\omega_{2}$ and normalizing space, and the wave function with $\omega_{0}^{-1}$, $l_{0}=\sqrt{\hbar / \left(m \omega_{0} \right)}$, and $1/\sqrt{l_{0}}$. GP equation becomes as follows:
\begin{equation}
\ri\hbar \frac{\partial }{\partial t}\Psi \left( x,t\right) =-\frac{\hbar ^{2}%
}{2m}\frac{\partial ^{2}}{\partial x^{2}}\Psi \left( x,t\right) +\left[
V\left( x\right) +g_\text{1D}\left\vert \Psi \left( x,t\right) \right\vert
^{2}\right] \Psi \left( x,t\right),  \label{eqn2.1}
\end{equation}

Here, interaction is reduced $g_\text{1D}=2\omega_{r} a_{s}/\left(\omega_{0} l_{0}\right)$ and the potential $V_\text{ext}$ is normalized by $\hbar \omega_{0}$. We use parameters $\omega_{r}=19$~Hz and $\omega_{0}=250$~Hz that are used in the experiments \cite{Work90}. From these parameters  $\sqrt{\hbar / \left(m \omega_{0} \right)} \approx 3.3$~$\muup$m for $^{23}$Na. The $s$-wave length is chosen $5-10$ nm \cite{Work91}, therefore, $g_\text{1D}$ could be from $0.011$ to $0.022$.

\section{Tilted bichromatic optical lattice potential}

\begin{figure}[!t]
\centerline{\includegraphics[width=0.45\textwidth]{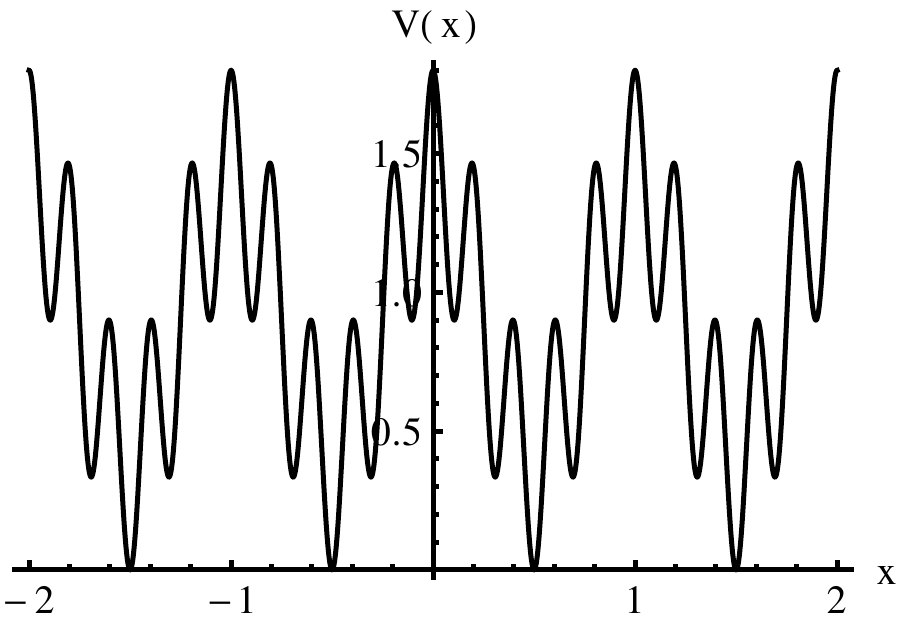}   \hspace{0.5cm}\includegraphics[width=0.45\textwidth]{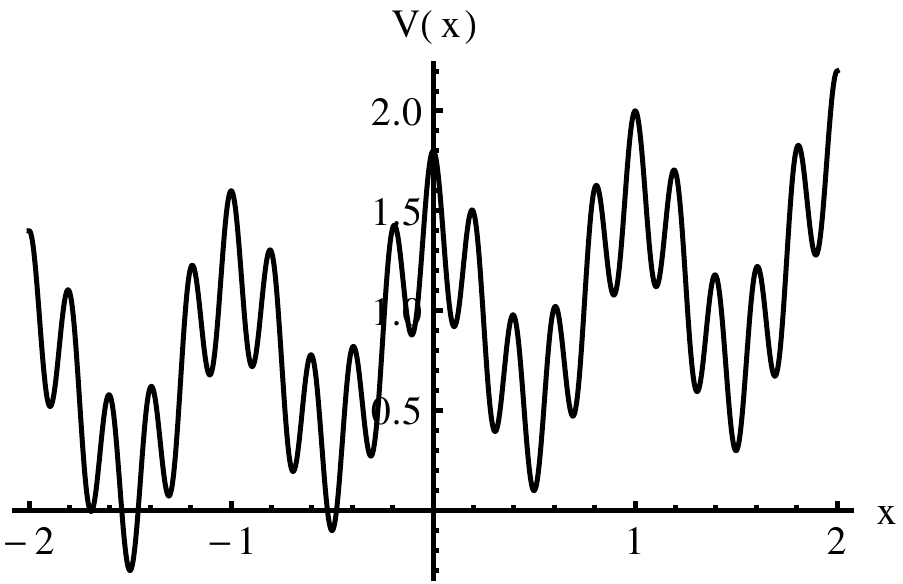}  }
\begin{center} (a) \hspace{6cm} (b) \end{center}
\vspace*{8pt}
\caption{Plot of the bichromatic optical lattice potential with the parameters $\nu_{1}=1, \nu_{2}=0.8, \omega_{1}=2\piup, \omega_{2}=5\piup$, (a) $\zeta=0$, (b) $\zeta=0.1$. }
\label{fig2}
\end{figure}

We choose the external trapping potential as follows:
\begin{equation}
V_\text{ext} \left(x \right )=V(x)+Fx\,, \label{eqn3.1}
\end{equation}
where the constant force $F$ describes $F=ma$. Here, $m$ is the mass of atoms and $a$ is the acceleration wich generates a tilted optical
potential as shown in figure~\ref{fig2}, accelerates the atoms in the $x$ direction with linearly increasing flow density and leads to the atoms tunnelling out of the traps \cite{Work25, Work26, Work34, Work29}. Moreover, $V(x)$ is the optical potential. In this study, the optical potential is given in the form of
\begin{equation}
V(x)=V_{1} \cos^{2} (\omega_{1}x)+V_{2} \cos^{2} (\omega_{2}x)\,, \label{eqn3.2}
\end{equation}
here, $V_{1}$ and $V_{2}$ are the respective amplitudes. In order to obtain a simple description and for a better understanding of the BEC dynamics, we consider only the solution of equation~(\ref{eqn2.1}) as below \cite{Work29,Work30}
\begin{equation}
\Psi \left( x,t\right) =\Phi \left( x\right) \re^{\frac{-\ri\mu t}{\hbar }},
\label{eqn3.3}
\end{equation}
here, $\mu$ is the chemical potential of the condensate and $\Phi(x)$  is a real function independent of time. $\Phi(x)$ is normalized to the total number of particles in the system, i.e.,
\begin{equation}
\int \left\vert \Phi \left( x\right) \right\vert ^{2}\rd x=N, \label{eqn3.4}
\end{equation}
where $N$ is the particle number. Substitution of equations~(\ref{eqn3.2}) and (\ref{eqn3.3})  into equation~(\ref{eqn2.1}) , yields
\begin{eqnarray}
\mu \Phi \left(x \right)&=&-\frac{\hbar ^{2}}{2m}\frac{\partial ^{2}}{\partial x^{2}}\Phi \left( x\right)  \nonumber\\
&+&\left[
V _{1}\cos^{2}\left( w_{1}x\right)+V_{2}\cos^{2}\left( w_{2}x\right) +Fx +g_\text{1D}\left\vert \Phi \left( x\right) \right\vert
^{2}\right] \Phi \left( x\right), \label{eqn3.5}
\end{eqnarray}

For simplicity, we rescale the wave function and present dimensionless parameters as $\upsilon _{1}={2mV_{1}}/{\hbar ^{2}}$, $\upsilon _{2}={2mV_{2}}/{\hbar ^{2}}$, $\gamma ={2m\mu }/{\hbar ^{2}}$, $\zeta={2mF}/{\hbar ^{2}}$, $\eta = {2mg_{0}}/{\hbar ^{2}}$.  The system equation can be written in the following form
\begin{equation}
\frac{\rd^2\Phi }{\rd x^2}=\left[ \upsilon _{1}\cos^{2}\left( w_{1}x\right) +\upsilon
_{2}\cos^{2}\left( w_{2}x\right) +\zeta x-\gamma +\eta \left\vert \Phi
\right\vert ^{2}\right] \Phi\,,  \label{eqn3.6}
\end{equation}

Inserting $\Phi(x)=\phi(x) \re^{\text{i}\theta(x)}$  into  (\ref{eqn3.6})   leads to two coupled equations which comes from separetly analyzing the real and the imaginary parts, 
\begin{subequations}
\begin{equation}
\frac{\rd^{2}\phi }{\rd x^{2}}=\phi \left( \frac{\rd\theta }{\rd x}\right) ^{2}+\left[
\upsilon _{1}\cos^{2}\left( w_{1}x\right) +\upsilon _{2}\cos^{2}\left(
w_{2}x\right) +\zeta x-\gamma +\eta \left\vert \phi \right\vert ^{2}\right]
\phi\,,  \label{eqn3.8a}
\end{equation}
\begin{equation}
 \frac{\rd^{2}\theta }{\rd x^{2}}+2\frac{1}{\phi}\frac{\rd\theta}{\rd x}\frac{\rd\phi}{\rd x}=0. \label{eqn3.8b}
\end{equation}
\end{subequations}
where $\phi(x)$ and $\theta(x)$ are real functions. The first derivative $2{\rd \theta }/{\rd x}$  in equation~(\ref{eqn3.8b}) shows the velocity and $\phi ^{2}=n$ is the number of density of atoms. Therefore, equation~(\ref{eqn3.8b}) denotes that there exist a steady current $J$ which is obtained from the first integration constant of equation~(\ref{eqn3.8b}) as below
\begin{equation}
J=2\phi^{2}\frac{\rd\theta}{\rd x}\,.\label{eqn3.9}
\end{equation}

Equation~(\ref{eqn3.9}) represents a steady superfluid in the system. If we put $J$ into equation~(\ref{eqn3.8a}), we have a nonlinear equation
\begin{equation}
\frac{\rd^{2}\phi }{\rd x^{2}}=\frac{J^{2}} {4\phi ^{3}}+\left[ \upsilon
_{1}\cos^{2}\left( w_{1}x\right) +\upsilon _{2}\cos^{2}\left( w_{2}x\right)
+\zeta x-\gamma +\eta \left\vert \phi \right\vert ^{2}\right] \phi .\label{eqn3.10}
\end{equation}

It is difficult to obtain the exact solution of eqaution~(\ref{eqn3.10})  because of its complexity. Thus, numerical solutions were performed. For numerical calculation, we reduce the system to the first-order by the transformation
\begin{subequations}
\begin{equation}
\frac{\rd x_{1}}{\rd x}=y_{1}\,, \label{eqn3.11a}
\end{equation}
\begin{equation}
\frac{\rd y_{1}}{\rd x}=\frac{J^{2}} {4x_{1} ^{3}}+\left[ \upsilon
_{1}\cos^{2}\left( w_{1}x\right) +\upsilon _{2}\cos^{2}\left( w_{2}x\right)
+\zeta x-\gamma +\eta \left\vert x_{1} \right\vert ^{2}\right] x_{1}. \label{eqn3.11b}
\end{equation}
\end{subequations}
where $\phi=x_{1}$ and ${\rd \phi}/{\rd x}=y_{1}$. The superposition of two laser beams of different frequencies can generate a bichromatic potential. Therefore, bichromatic potential can be referred to as double well potential. In this study, we add an extra tilted force to the bichromatic potential. This causes an acceleration of bosons in the lattice potential. Bosons can be allowed to jump from one trap to another. 

Recently, the regular and chaotic solutions of the BEC system in 1D tilted bichromatical optical lattice potential have been investigated by constructing its Poincare sections in phase space \cite{Work31}. Moreover, the dynamics of a Bose-Einstein condensate system under the Gaussian white noise has been studied \cite{Work32}.

\section{Chaos synchronization in BEC}

Chaos synchronization means making two (or more) systems oscillate in the same manner by using the active control parameters \cite{Work70}. Master-slave control technique is the most frequently used synchronization approach used for the chaotic systems. In the master-slave scheme, the behaviors of the slave system are controlled with the operation of a master system \cite{Work80}.
The  master and slave systems are defined as follows:
\begin{subequations}
\begin{equation}
\frac{\rd x_{1}}{\rd x}=y_{1}\,, \label{eqn4.1a}
\end{equation}
\begin{equation}
\frac{\rd y_{1}}{\rd x}=\frac{J^{2}} {4x_{1} ^{3}}+\left[ \upsilon
_{1}\cos^{2}\left( w_{1}x\right) +\upsilon _{2}\cos^{2}\left( w_{2}x\right)
+\zeta x-\gamma +\eta \left\vert x_{1} \right\vert ^{2}\right] x_{1}\,,  \label{eqn4.1b}
\end{equation}
\end{subequations}
and 
\begin{subequations}
\begin{equation}
\frac{\rd x_{2}}{\rd x}=y_{2}+u_{1}(x)\,, \label{eqn4.2a}
\end{equation}
\begin{equation}
\frac{\rd y_{2}}{\rd x}=\frac{J^{2}} {4x_{2} ^{3}}+\left[ \upsilon
_{1}\cos^{2}\left( w_{1}x\right) +\upsilon _{2}\cos^{2}\left( w_{2}x\right)
+\zeta x-\gamma +\eta \left\vert x_{2} \right\vert ^{2}\right] x_{2}+u_{2}(x)\,,   \label{eqn4.2b}
\end{equation}
\end{subequations}
where $\phi= x_{2}$, ${\rd \phi}/{\rd x}=y_{2}$, $u_{1}$ and $u_{2}$  are the nonlinear controllers, so that two chaotic systems can be synchronized. In this case, the error dynamics are $e_{1}=x_{2}-x_{1}$ and $e_{2}=y_{2}-y_{1}$. The error dynamics is determined by subtracting equations~(\ref{eqn4.1a}), (\ref{eqn4.1b}) from equation~(\ref{eqn4.2a}), (\ref{eqn4.2b})
\begin{subequations}
\begin{gather}
\frac{\rd e_{1}}{\rd  x}=e_{2}+u_{1}(x)\,,  \label{eqn4.3a}
\end{gather}
\begin{eqnarray}
\frac{\rd e_{2}}{\rd x}&=&\frac{J^{2}} {4x_{2} ^{3}}-\frac{J^{2}} {4x_{1} ^{3}} + \left[ \upsilon
_{1}\cos^{2}\left( w_{1}x\right) +\upsilon _{2}\cos^{2}\left( w_{2}x\right)
+\zeta x-\gamma +\eta \left\vert x_{2} \right\vert ^{2}\right]e_{1}+u_{2}(x).  \label{eqn4.3b}
\end{eqnarray}
\end{subequations}

We write the error functions as below,
\begin{subequations}
\begin{equation}
u_{1}(x)=\tau_{1}(x)\,,  \label{eqn4.4a}
\end{equation}
\begin{equation}
u_{2}(x)=-\frac{J^{2}} {4x_{2} ^{3}}+\frac{J^{2}} {4x_{1} ^{3}}-\eta (x_{2}^{2}-x_{1}^{2})+\tau_{2}(x). \label{eqn4.4b}
\end{equation}
\end{subequations}

Substituting control functions into equation~(\ref{eqn4.5a})  and (\ref{eqn4.5b}), the error system becomes,
\begin{subequations}
\begin{align}
\frac{\rd e_{1}}{\rd x}&=e_{2}+\tau_{1}(x)\,, \label{eqn4.5a}\\
\frac{\rd e_{2}}{\rd x}&=\left[ \upsilon
_{1}\cos^{2}\left( w_{1}x\right) +\upsilon _{2}\cos^{2}\left( w_{2}x\right)
+\zeta x-\gamma +\eta \right] e_{1}+\tau_{2}(x). \label{eqn4.5b}
\end{align}
\end{subequations}

The system can be controlled with control inputs $\tau_{1}$ and $\tau_{2}$ as function of $e_{1}$ and $e_{2}$. These feedbacks stabilize the system where $\lim_{x\to\infty} ||e(x)||=0$.  We choose control inputs $\tau_{1}$ and $\tau_{2}$ as follows:
\begin{equation}
\begin{bmatrix}
\tau_{1} \\
\tau_{2}
\end{bmatrix}
=D
\begin{bmatrix}
e_{1} \\
e_{2}
\end{bmatrix}, \label{eqn4.6}
\end{equation}
where $D=
\begin{bmatrix}
a & b \\
c & d
\end{bmatrix}$
is a $2 \times 2$ constant feedback matrix. The system can be written as follows:
\begin{equation}
\begin{bmatrix}
\frac{\rd e_{1}}{\rd x} \\
\frac{\rd e_{2}}{\rd x}
\end{bmatrix}
=C
\begin{bmatrix}
e_{1} \\
e_{2}
\end{bmatrix}, \label{eqn4.7}
\end{equation}
where $C$ is is the coefficient matrix as below
\begin{equation}
C=
\begin{bmatrix}
0+a & 1+b \\
\upsilon
_{1}\cos^{2}\left( w_{1}x\right) +\upsilon _{2}\cos^{2}\left( w_{2}x\right)
+\zeta x-\gamma +\eta +c & 0+d
\end{bmatrix}. \label{eqn4.8}
\end{equation}
 
Based on the Routh-Hurwitz criterion \cite{Work81}, Lyapunov stability theory and active control strategy, two identical BEC are synchronized. If we choose  $a=-1$, $b=-1$, $d=-1$, \newline
 $c=-\left[\upsilon_{1}\cos^{2}\left( w_{1}x\right) +\upsilon _{2}\cos^{2}\left( w_{2}x\right)+\zeta x-\gamma +\eta \right]$, 
\begin{subequations}
\begin{equation}
\tau_{1}(x)=-e_{1}-e_{2}\,, \label{eqn4.9a}
\end{equation}
\begin{equation}
 \tau_{2}(x)=-\left[\upsilon
_{1}\cos^{2}\left( w_{1}x\right) +\upsilon _{2}\cos^{2}\left( w_{2}x\right)
+\zeta x-\gamma +\eta \right]e_{1}-e_{2}\,.  \label{eqn4.9b}
\end{equation}
\end{subequations}

Finally we get a slave system as below,
\begin{subequations}
\begin{equation}
\frac{\rd x_{2}}{\rd x}=y_{2}+(-e_{1}-e_{2})\,, \label{eqn4.10a}
\end{equation}
\begin{equation}
\begin{split}
\frac{\rd y_{2}}{\rd x}&=\frac{J^{2}} {4x_{1} ^{3}}+\left[ \upsilon
_{1}\cos^{2}\left( w_{1}x\right) +\upsilon _{2}\cos^{2}\left( w_{2}x\right)
+\zeta x-\gamma +\eta \left\vert x_{2} \right\vert ^{2}\right] x_{2} \\&-\eta(x_{2}^{3}-x_{1}^3)-\left[\upsilon
_{1}\cos^{2}\left( w_{1}x\right) +\upsilon _{2}\cos^{2}\left( w_{2}x\right)
+\zeta x-\gamma +\eta \right]e_{1}-e_{2}\,.
\end{split} \label{eqn4.10b}
\end{equation}
\end{subequations}

\section{Numerical simulations}
In this section, we investigate the simulation results for chaotic synchronization of the systems~(\ref{eqn4.1a}), (\ref{eqn4.1b}) and (\ref{eqn4.2a}), (\ref{eqn4.2b}). In the numerical simulations, we have written a Maxima code that applies the fourth order Runge Kutta algorithm which is used to solve the sets of differential equations related to the master and slave systems with step size 0.1 and step length 3500. We selected the parameters of the BEC system as $\zeta=0.1$, ${J}=0.4$, $ v _{1}=1,v _{2}=0.8,w_{1}=2\piup, w_{2}=5\piup ,\gamma =0.5,\eta =-0.015$, respectively. Shock wave dynamics can be found in our recently published paper for these parameters \cite{Work31}.

\begin{figure}[!t]
\centering{\begin{tabular}{c c}
\includegraphics[width=0.45\textwidth]{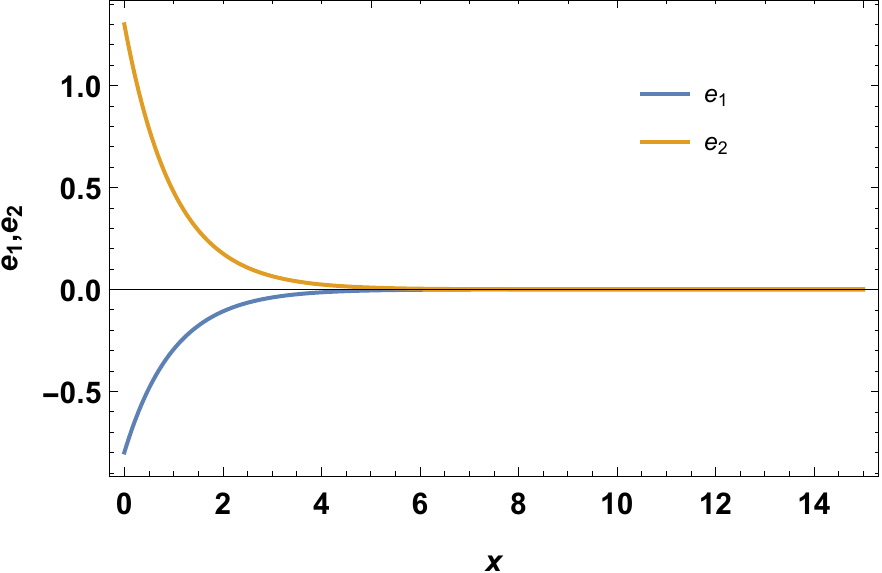}
&
\includegraphics[width=0.45\textwidth]{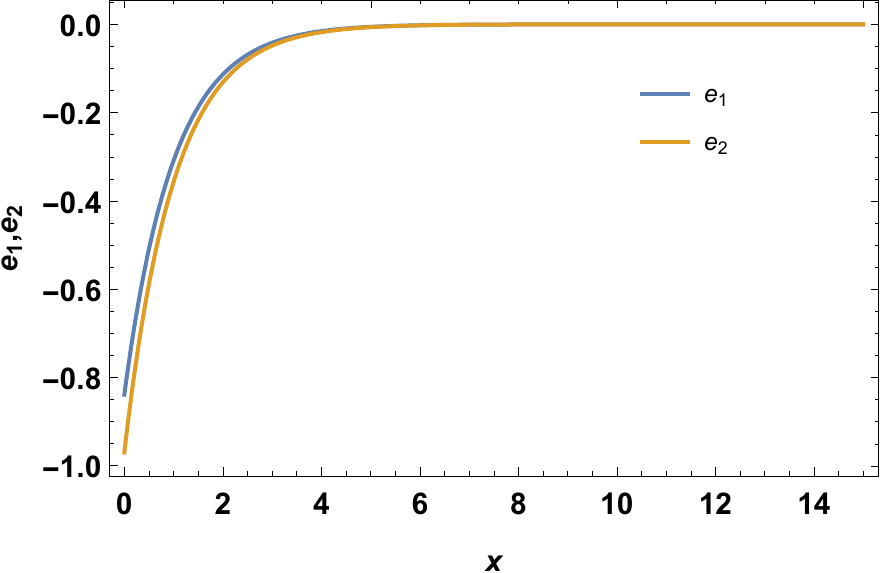}   \\
(a) & (b)

\end{tabular}} \\

\vspace*{8pt}
\caption{(Colour online) Spatial evolution of $(e_{1},e_{2})$ with the controller activated at the $x=0$ for (a) $(e_{1}(0),e_{2}(0))=(-0.8,1.3)$ and (b)  $(e_{1}(0),e_{2}(0))=(-0.839,-0.97)$.}
\label{fig3}
\end{figure}

\begin{figure}[!t]
\begin{tabular}{c c}
\includegraphics[width=0.45\textwidth]{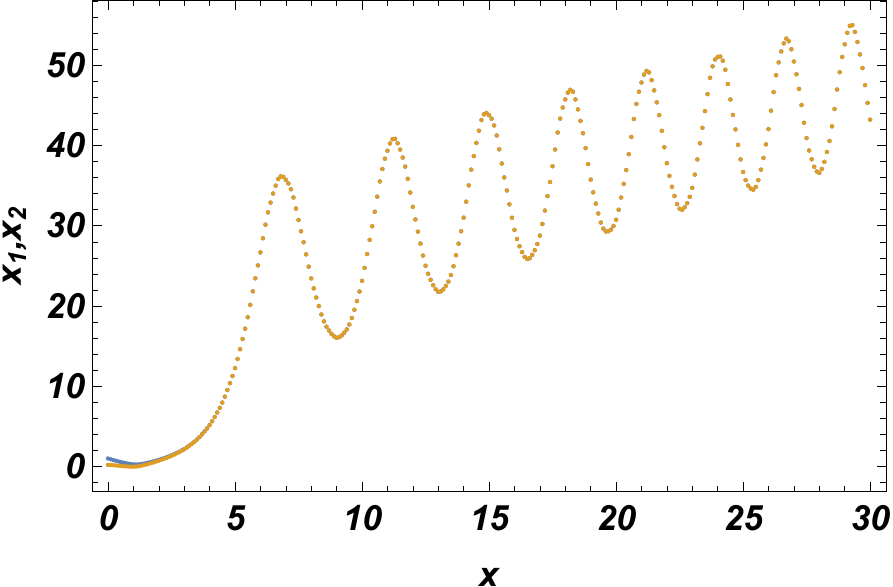}  
&
\includegraphics[width=0.45\textwidth]{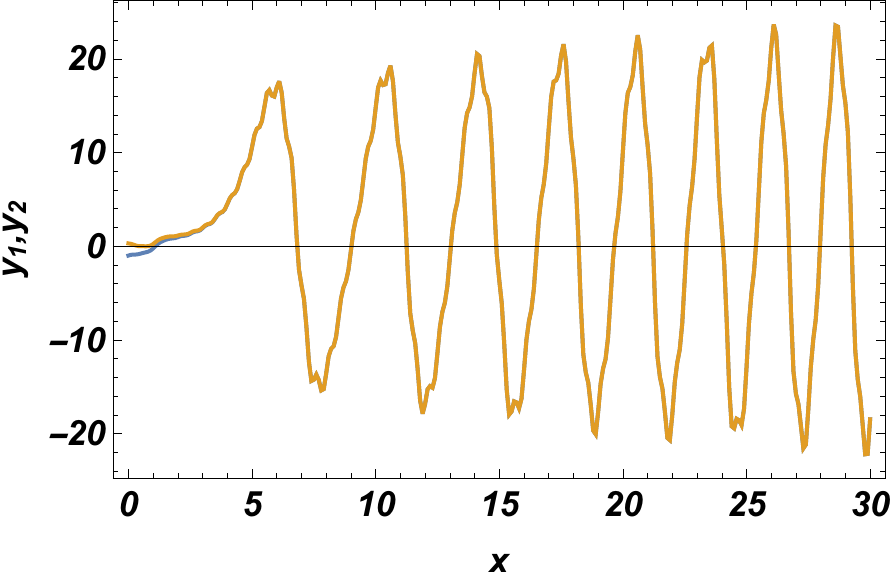}  
 \end{tabular} \\
 \centering{(a) \hspace{165pt} (b)}
 \vspace*{8pt}
 
\begin{tabular}{c c}
 \includegraphics[width=0.45\textwidth]{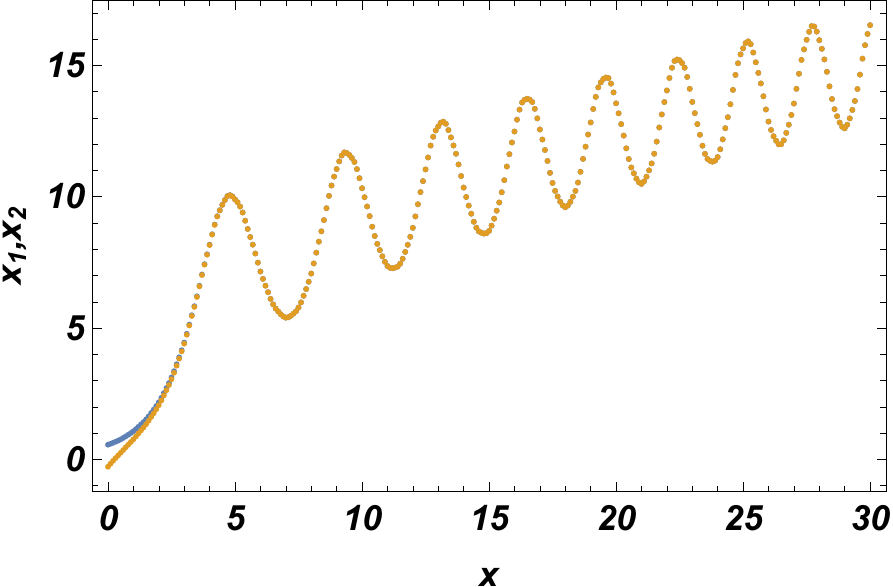}  
&
\includegraphics[width=0.45\textwidth]{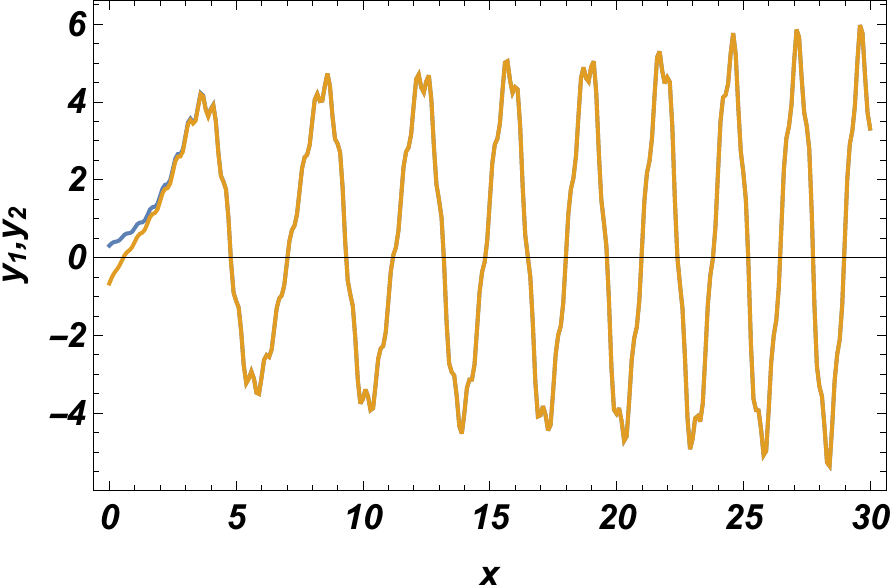}   \end{tabular} \\
\centering{(c) \hspace{165pt} (d)}

\vspace*{8pt}
\caption{(Colour online) Synchronized  master-slave systems using the active control method for $0 \leq x \leq 30$,  $\zeta=0.1$, ${J}=0.4$, $ v _{1}=1,v _{2}=0.8,w_{1}=2\piup ,w_{2}=5\piup ,\gamma =0.5,\eta =-0.015$  (a) $\left(x_{1}(0),x_{2}(0)\right)=\left(1,0.2\right)$,   (b)~$\left(y_{1}(0),y_{2}(0)\right)=\left(-1,0.3\right)$, (c)  $\left(x_{1}(0),x_{2}(0)\right)=\left(0.56,0.279\right)$,  (d) $\left(y_{1}(0),y_{2}(0)\right)=\left(0.3,0.67\right)$. }
\label{fig4}
\end{figure}

\begin{figure}[!t]
\begin{tabular}{c c}
\includegraphics[width=0.44\textwidth]{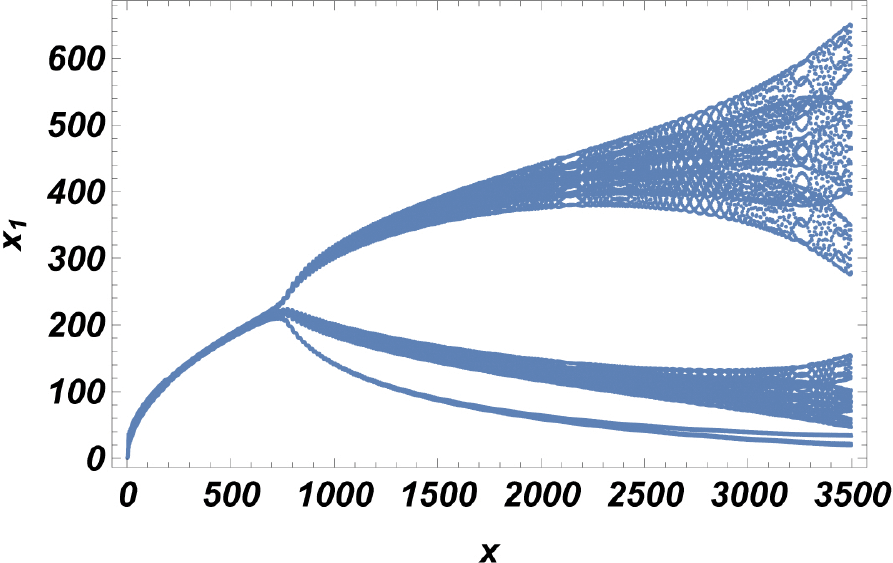}  
&
\includegraphics[width=0.44\textwidth]{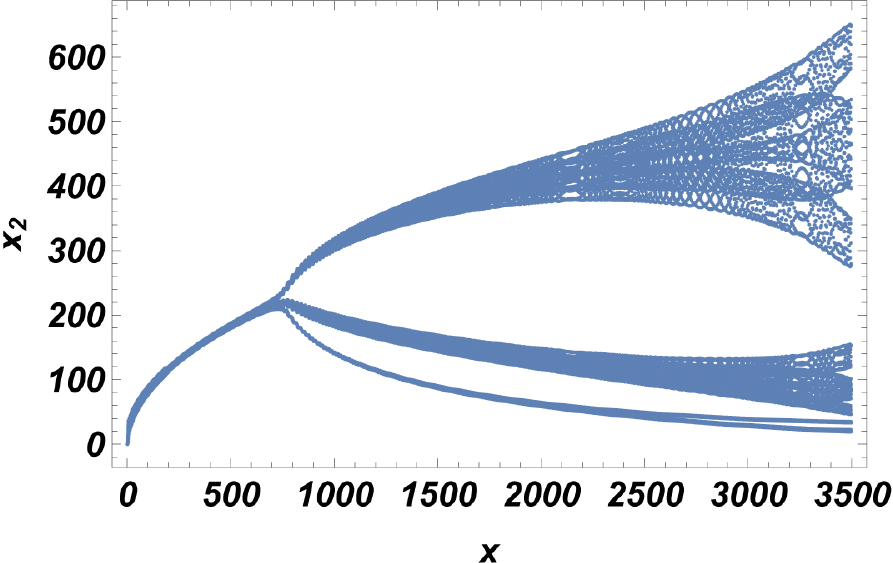}  
 \end{tabular} \\
 \centering{(a) \hspace{165pt} (b)}
 \vspace*{8pt}
 
 \begin{tabular}{c c}
 \includegraphics[width=0.44\textwidth]{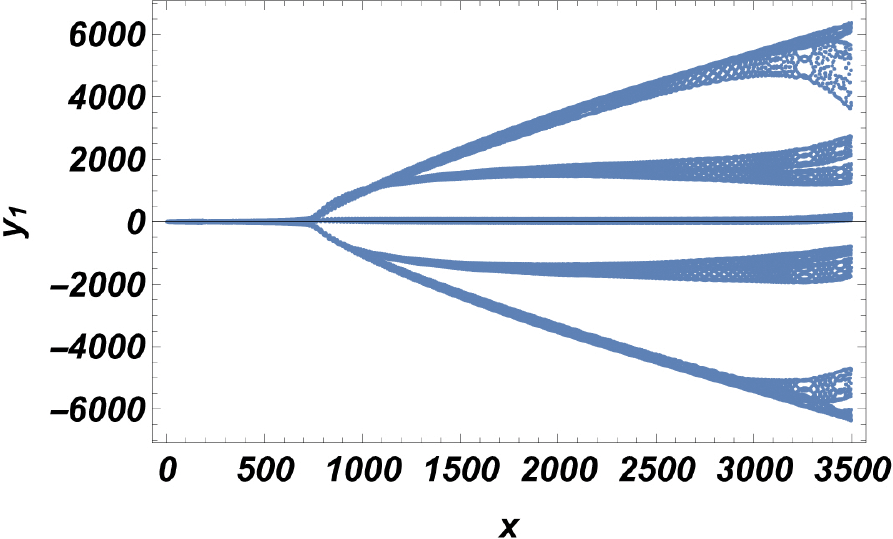}  
&
\includegraphics[width=0.44\textwidth]{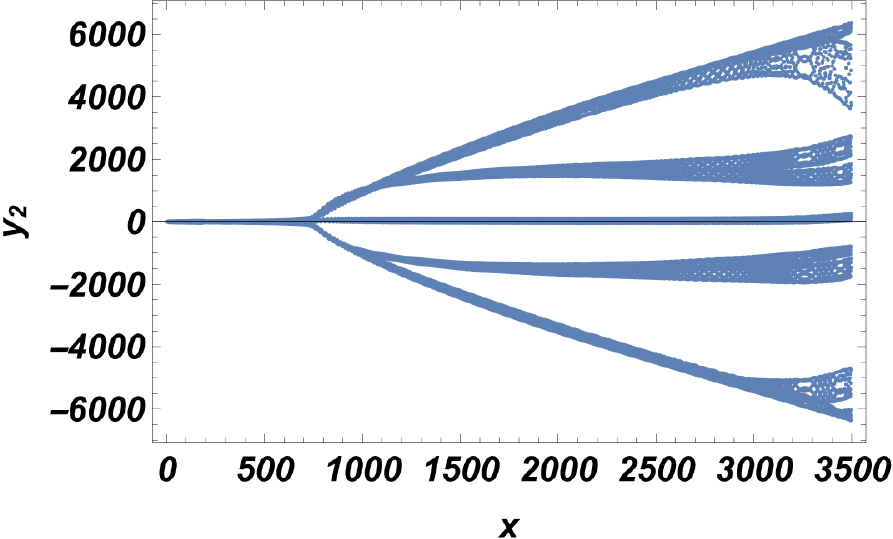}   
\end{tabular} \\
\centering{(c) \hspace{165pt} (d)}
 \vspace*{8pt}
\vspace{-4mm}
\caption{(Colour online) Chaotic synchronization the master-slave systems using active control method for  $\zeta=0.1$, ${J}=0.4$, $ v _{1}=1, v _{2}=0.8, w_{1}=2\piup, w_{2}=5\piup,\gamma =0.5,\eta =-0.015$ and the initial conditions for master systems  (a) $x_{1}(0)=1$  and (c)  $y_{1}(0)=-1$, slave systems (b) $x_{2}(0)=0.2$ and (d) $y_{2}(0)=0.3$. }
\label{fig5}
\end{figure}

\begin{figure}[!t]
\begin{tabular}{c c}
\includegraphics[width=0.42\textwidth]{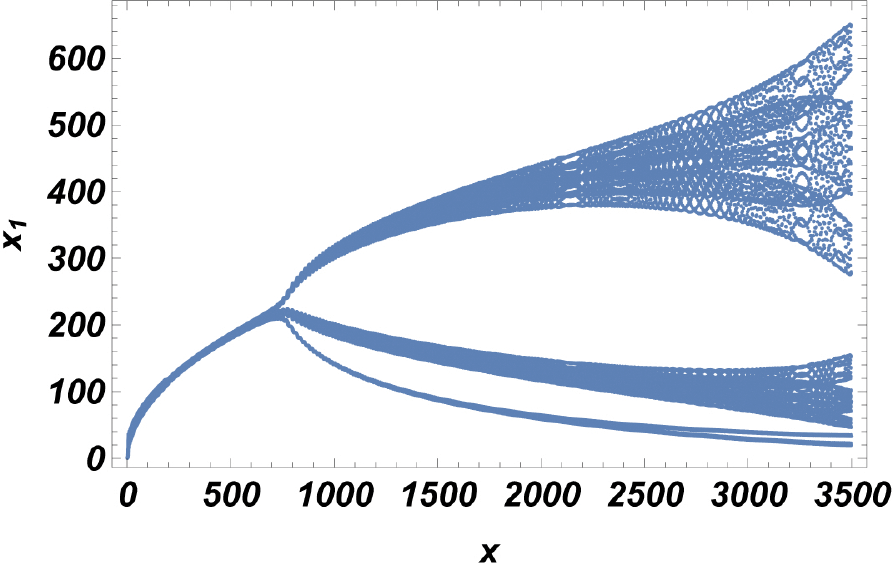}  
&
\includegraphics[width=0.42\textwidth]{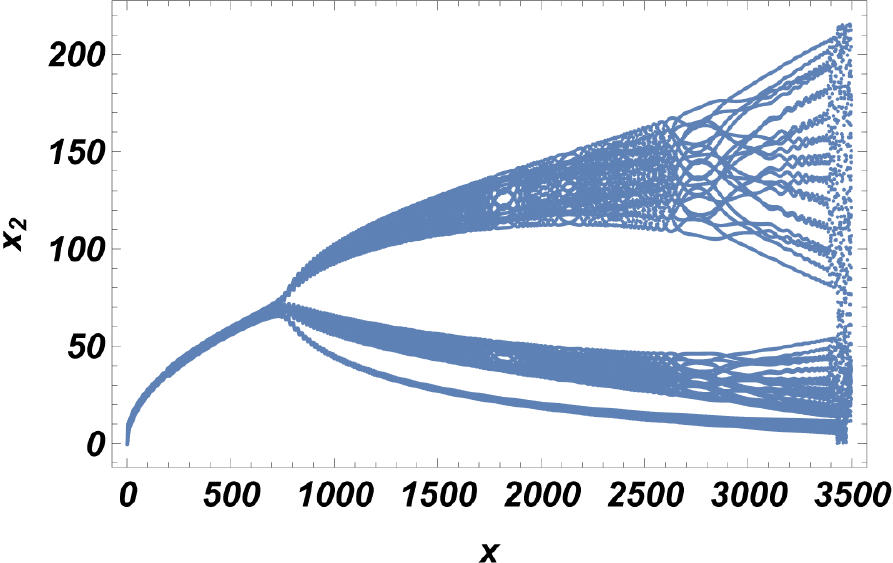}  
 \end{tabular} \\
 \centering{(a) \hspace{165pt} (b)}
 \vspace*{8pt}
 \vspace{-2mm}
\begin{tabular}{c c}
 \includegraphics[width=0.44\textwidth]{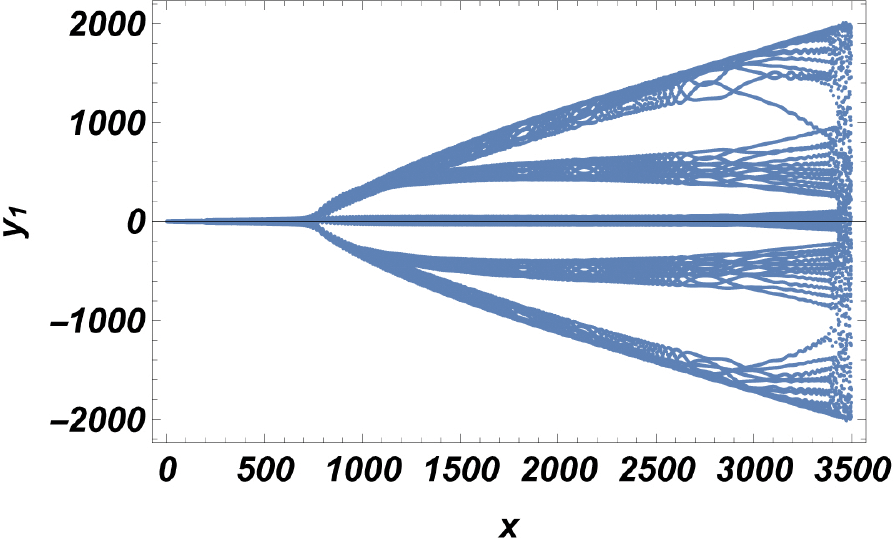}  
&
\includegraphics[width=0.44\textwidth]{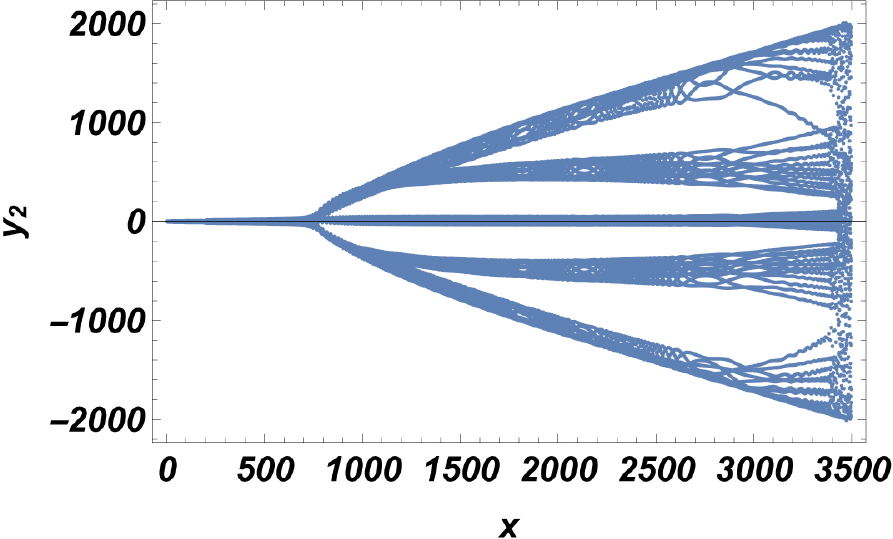}   \end{tabular} \\
\centering{(c) \hspace{165pt} (d)}

\vspace*{8pt}
\vspace{-5mm}
\caption{(Colour online) Chaotic synchronization the master-slave systems using active control method for  $\zeta=0.1$, ${J}=0.4$, $ v _{1}=1, v _{2}=0.8, w_{1}=2\piup, w_{2}=5\piup ,\gamma =0.5,\eta =-0.015$ and the initial conditions for master systems (a) $x_{1}(0)=0.56$  and (c)  $y_{1}(0)=0.3$, slave systems (b) $x_{2}(0)=0.279$ and (d) $y_{2}(0)=0.67$.}
\label{fig6}
\end{figure}

For the numerical calculations, we choose the arbitrary possible intial condtions between $-1$ and $1$. In the chaotic regime, the velocity field and atomic density may not be determined exactly in experiment due to fluctuations of the atomic thermal cloud \cite{Work1a, Work1b}. Thus, we choose possible intial conditions randomly for numerical calculations. In figure~\ref{fig3}  we plot the error functions from equation~(\ref{eqn4.5a}) and (\ref{eqn4.5b}) for two different initial conditions (a) $(e_{1}(0),e_{2}(0))=(-0.8,1.3)$ and (b)  $(e_{1}(0),e_{2}(0))=(-0.839,-0.97)$, respectively. Control functions start at $x=0$. As a result, the system rapidly tends to synchronization. Figures~\ref{fig4}, \ref{fig5} and \ref{fig6} represent the solution of master and slave systems. In figure~\ref{fig4}  spatial evolutions with the controller are given for the  initial conditions (a) $\left(x_{1}(0),x_{2}(0)\right)=\left(1,0.2\right)$,   (b) $\left(y_{1}(0),y_{2}(0)\right)=\left(-1,0.3\right)$, (c)  $\left(x_{1}(0),x_{2}(0)\right)=\left(0.56,0.279\right)$,  (d) $\left(y_{1}(0),y_{2}(0)\right)=\left(0.3,0.67\right)$ and $x_\text{max}=30$. We solve the master and slave system from $x=0$ to $x_\text{max}=3500$ with the step size $0.1$ in figures~\ref{fig5} and \ref{fig6}. According to spatial evolution, the system shows chaotic dispersive shock-wave-like dynamics in the phase space~\cite{Work31}. In addition, the system is synchronized along the flow.

\pagebreak
\newpage

\section{Conclusion}

It is known that the chaos in BEC systems can undermine the stability of the condensates playing a destructive role \cite{Work34}. Thus, the studies on predicting and controlling the chaos on BEC systems are quite important. An active control method can be used in time and space dependent systems. Recently, synchronization of BEC via an active control method was investigated for the time dependence~\cite{Work1c, Work1d}.In this study, we have studied the synchronization of the BEC system held in a 1D tilted bichromatical optical lattice potential via the master-slave active control technique for real space dependence. We first present the system under 1D tilted bichromatical optical lattice potential, and then we investigate chaotic synchronization in BEC. The chaotic synchronization system consists of the master and slave system. The control functions provide the situation when the states of the chaotic slave system exponentially synchronize with the state of the master system with different initial conditions. The initial values of the master-slave system are taken as  $(x_{1}[0],y_{1}[0])=(1,-1)$ and  $(x_{2}[0], y_{2}[0])=(0.2,0.3)$  and $(x_{1}[0],y_{1}[0])=(0.56,0.3)$ and  $(x_{2}[0], y_{2}[0])=(-0.279,-0.67)$, respectively, for numerical simulation. The synchronization errors rapidly converge to zero. From the physical point of view, the technique used is based on the  Routh-Hurwitz criterion and is quite useful in the ultra-cold quantum superfluid atoms. Numerical results verify the accuracy of the applied method, and the obtained results show that two idential BEC could be synchronized by an active control method for real space dependence.

\ukrainianpart

\title{Синхронізація типу керівний-керований конденсату Бозе-Ейнштейна  в  1D похилій біхроматичній оптичній гратці}
\author{E. Тосіалі\refaddr{label1}, Ф. Айдогмус\refaddr{label2}}
\addresses{
	\addr{label1} Стамбульський університет Білгі, професійне училище медичних послуг,  34387 Стамбул/Шішлі, Туреччина
	\addr{label2} Стамбульський університет, фізичний факультет, 34134 Стамбул, Туреччина}

\makeukrtitle

\begin{abstract}
	У статті досліджується синхронізація хаотичної поведінки у моделі  конденсату Бозе-Ейнштейна, поміщеного у потенціал  1D похилої біхроматичної оптичної гратки з використанням методу активного контролю. Синхронізація представлена конфігурацією  ``керівний-керований'', яка передбачає, що керівна система  розвивається вільно та запускає динаміку керованої системи. Представлено числові симуляції, що підтверджують практичність та ефективність застосованих контролерів.
	\keywords синхронізація, конденсат Бозе-Ейнштейна, рівняння Гросса-Пітаєвського, оптичний гратковий потенціал
\end{abstract}

\end{document}